\documentclass[prd,aps]{revtex4}
\usepackage{graphicx}

\begin{document}
\title{ A relational solution to the problem of time in quantum
mechanics and quantum gravity induces a fundamental mechanism for
quantum decoherence}

\author{Rodolfo Gambini$^{1}$, Rafael A. Porto$^{2}$ and Jorge Pullin$^{3}$}
\affiliation {1. Instituto de F\'{\i}sica, Facultad de Ciencias,
Igu\'a 4225, esq. Mataojo, Montevideo, Uruguay. \\ 
2. Department of Physics, Carnegie Mellon University, Pittsburgh PA 15213\\
3. Department
of Physics and Astronomy, Louisiana State University,
Baton Rouge, LA 70803-4001}
\date{February 22nd 2004}

\begin{abstract}
The use of a relational time in quantum mechanics is a framework in
which one promotes to quantum operators all variables in a system, and
later chooses one of the variables to operate like a
``clock''. Conditional probabilities are computed for variables of the
system to take certain values when the ``clock'' specifies a certain
time. This framework is attractive in contexts where the assumption of
usual quantum mechanics of the existence of an external, perfectly
classical clock, appears unnatural, as in quantum cosmology. Until
recently, there were problems with such constructions in ordinary
quantum mechanics with additional difficulties in the context of
constrained theories like general relativity. A scheme we recently
introduced to consistently discretize general relativity removed such
obstacles. Since the clock is now an object subject to quantum
fluctuations, the resulting evolution in the time is not exactly
unitary and pure states decohere into mixed states. Here we work out
in detail the type of decoherence generated, and we find it to be of
Lindblad type. This is attractive since it implies that one can have
loss of coherence without violating the conservation of energy. We
apply the framework to a simple cosmological model to illustrate how a
quantitative estimate of the effect could be computed. For most
quantum systems it appears to be too small to be observed, although
certain macroscopic quantum systems could in the future provide a
testing ground for experimental observation.
\end{abstract}

\maketitle

\section{Introduction}
\subsection{Relational time in quantum mechanics}
There exists interest in applying the rules of quantum mechanics in situations
where there does not exist an obvious way of considering an external perfectly
classical clock. An example is provided by quantum cosmology, presumably the
correct description of the universe close to the big bang. Sometimes
the discussion of quantum cosmology gets further complicated by the
problems associated with the quantization of general relativity. But
it should be emphasized that the quantization remains troublesome even
if one considers simplified model cosmologies, for instance described
by Newton's theory (an example is the Barbour--Bertotti model \cite{BaBe}).

Given the lack of an external time, one could try to use a variable
internal to the system under study as a clock. Such a variable could
not play the role of ``t'' in a Schr\"odinger equation, since one
expects it to be on a similar footing as the other variables in the
problem and therefore should be subject to quantum fluctuations. There
have been proposals in the past to build a quantum mechanics using an
internal variable as a clock. We will call these proposals
``relational time'' (see the next subsection for clarifications on 
terminology). What one does is to promote all variables in the
theory to quantum operators, then choose from among the variables one
(or several) that will operate as a ``clock'' and then compute
conditional probabilities for the other variables to take certain
values when the ``clock'' variable takes a given value. If one now
considers the system as possessing an external perfect classical clock
and builds an ordinary Schr\"odinger description for the system, and
if the variable chosen as internal clock behaves semiclassically in 
such a description, then the relational picture and the Schr\"odinger
picture will give similar descriptions of the system. If the variable 
chosen as ``clock'' has large quantum fluctuations, then both
descriptions will differ. It is clear that since the ``internal
clock'' will always have some quantum fluctuations, that both
descriptions can never be identical. It is therefore of interest to
address by how much both descriptions differ. This is one of the main
purposes of this paper.

In ordinary quantum mechanics, there is an old argument due to Pauli
that one cannot promote a dynamical variable to a quantum operator
associated with time, since this operator would have to be canonically
conjugate to the Hamiltonian, and since time is expected to be monotonous,
this would imply that the Hamiltonian cannot be bounded from below. We 
will see that in our approach this difficulty is not present.

There is an additional complication. If one considers a system as
alone in the universe, without external observers, there appear
constraints among the variables of the system. Such systems are called
``generally covariant'' or ``totally constrained''. For instance, in
the Barbour--Bertotti \cite{BaBe} model, which consists of a few
Newtonian particles alone in the universe, one has the constraint that
the total momentum, energy and the total angular momentum must have
fixed values. In such systems, although one can formally write them as
dependent on a parameter called ``t'', one can always reparameterize
them and there is no natural choice of a variable to identify with a
physical time. Sometimes this leads to calling them
``reparametrization invariant'' systems. In a cosmology based on
general relativity, one has the Hamiltonian and momentum
constraints. If one formulates the theory canonically in order to
quantize it, the constraints have to have vanishing Poisson brackets
with any quantity that can be considered physically. This presents
problems, since it implies that these quantities are constants of the
motion, and cannot be used as ``clocks''. Page and Wootters
\cite{PaWo} attempt to bypass this by choosing to build the relational
framework in terms of quantities that do not have vanishing Poisson
brackets with the constraints (that is, they choose to work at the
kinematical level.)  However, quantum mechanically, the states that
are annihilated by the constraints are distributional within the space
of kinematical states and do not lead to a good probabilistic
interpretation. The resulting propagators are proportional to the
Delta function and therefore ``they don't propagate'' as discussed in
detail by Kucha\v{r} \cite{Ku}.

We therefore see that the presence of constraints interfere with the
idea of attempting the introduction of a relational time in quantum
mechanics. One possibility to circumvent the problem is to get rid of
the constraints. We have recently proposed a discretization of field
theories that has the property that when applied to totally
constrained systems, the resulting discrete theories are constraint
free. In spite of this, they manage in certain circumstances, to
approximate well the continuum theories from which they were
generated. This allows the introduction of a relational notion of time
in the discrete theories. In this paper we would like to discuss, 
within the context of this approach, what are the quantitative effects
of evolving a system using the relational time.

It is clear that the relational time can be introduced in any context
that would produce a system that is free of constraints. The
discretized approach we consider is an example of such a context, but
there could exist others. 

\subsection{A note on terminology}

The idea of using a given dynamical variable in a closed
system as a clock is quite old. One can find traces of it in 
the work of Leibnitz, Mach, Bergmann, DeWitt, Page, Wootters,
Halliwell,
Ashtekar, Rovelli, Smolin, Mermin  and many others. For a partial list
see \cite{relational,Ku,PaWo}. The word ``relational'' is also used in
an attractive proposal of Rovelli \cite{RoRo} in which observables
(that is, quantities that have vanishing Poisson brackets with 
the constraints in constrained systems) are constructed via a
relational technique. The proposal differs from the one presented in
this paper. In it, the observables are constructed as dependent on 
a clock variable that is not promoted to a quantum operator like we 
do here.

\section{Consistent discretizations: brief summary}

\subsection{General framework}

We summarize here the consistent discretization approach, for a
lengthier discussion see \cite{DiGaPo}. Readers whose main interest is
not the details of how to handle general relativity can skip this
section. We start by considering a system with an action as in
ordinary classical mechanics dependent on a finite number of
configuration variables $q^a$, $S=\int d \tau
L(q^a,\dot{q}^a)$. Although this may appear oversimplificatory, recall
that when one discretizes a field theory one is really dealing with a
mechanical system. This Lagrangian should be considered as a very
general Lagrangian, in the sense that it can accommodate first order
formulations (in which case the momenta are considered as $q$'s) and
systems with explicit constraints in the Lagrangian (in which one
treats the Lagrange multipliers as $q$'s such that no $\dot{q}$
appears in the Lagrangian). The formalism can accommodate theories
invariant under reparameterizations, like general relativity, for that
reason we refer to the evolution parameter as $\tau$.

We now discretize the evolution parameter (in the case of a field
theory we also discretize space), and replace the derivatives via
$\dot{q}^a=(q_{n+1}^a-q_n^a)/\Delta \tau$ (other discretizations of the
derivatives are possible, but would require reworking the formulation).
The action will then become a sum $S=\sum_{i=0}^N L(q_n,q_{n+1})$ where
we have absorbed the interval of discretization into the definition of
$L(q,q_{n+1})$. The Lagrange equations are obtained by minimizing the
action with respect to each of the $q_n^b$,
\begin{equation}
{\partial L(q_n^a,q_{n+1}^a) \over \partial q_n^b} + {\partial
L(q_{n-1}^a,q_n^a) \over q_n^b} =0.
\end{equation}

We next introduce a type 1 canonical transformation that corresponds to the
evolution of the system from $n$ to $n+1$ with generating functional 
$-L(q_n^a,q_{n+1}^a)$,
\begin{eqnarray}
p_{n+1}^b &=& {\partial L(q_n^a,q_{n+1}^a) \over \partial q_{n+1}^a},\\
p_{n}^b &=& -{\partial L(q_n^a,q_{n+1}^a) \over \partial q_{n}^a}.
\end{eqnarray}
It is immediate to see, by evaluating the top equation at $n=n-1$
that the two equations are equivalent to the Lagrange equations. 
These two equations should be seen as an implicit  map between
$q_n^a,p_n^a$ and $q_{n+1}^a,p_{n+1}^a$. If the implicit map is invertible,
then the resulting transformation is indeed canonical, by construction.

If the map is not invertible, that is, if the determinant ${\rm det}
{\partial^2 L \over \partial q_n^a q_{n+1}^b}$ vanishes identically, one has
constraints. We denote them generically as $\phi^A(q_n^a,p_n^a)$,
$A=1\ldots M$.
The variables at instant $n+1$ are not completely determined by the
variables at instant $n$, and will depend on some free parameters
$V^B$, $B=1\ldots M$,
\begin{eqnarray}
q_{n+1} &=&q_{n+1}(q_n,p_n,V^B)\\
p_{n+1} &=&p_{n+1}(q_n,p_n,V^B).
\end{eqnarray}
It should be noted that the constraints not only occur at level $n$
but at all levels. We will treat them by considering the constraints 
at level $n$ and then ensuring conservation upon evolution. This 
can lead to four different situations: 1) imposing the constraints 
at all levels leads to the determination of some of the indeterminate 
parameters $V^B$; 2) new (secondary)  constraints appear; 3) 
the constraints  are preserved automatically; 4) the system is
incompatible. In the three compatible situations one generically 
ends up with evolution equations, potentially depending on some 
free parameters and potentially with extra constraints. It can be
shown that the resulting evolution equations \cite{DiGaPo} preserve
the Poisson brackets and (weakly) the constraint surface.

It should be noted that preserving constraints in the discrete theory
is harder to accomplish than in the continuum theory. In the latter
one only needs to show conservation infinitesimally, and the
calculations neglect higher order terms in the (infinitesimal) time
interval. In the discrete theory, since the evolution is finite, one
needs to show conservation preserving all powers of the time interval.
For this reason, in many circumstances constraints that are first
class in the continuum become second class when discretized. 
In our approach, constraints are eliminated by solving them for the Lagrange
multipliers. One therefore ends up with a discrete theory that is
free of constraints that nevertheless approximates a continuum theory
that has constraints. In the next subsection we illustrate this
procedure in detail by considering an example in general relativity.

To quantize the theory one has to deal with the constraints that may
be left, since as we mentioned it may not be always possible to solve
the constraints by choosing the Lagrange multipliers. If the remaining
constraints are second class one imposes them strongly and one
replaces Poisson brackets by Dirac brackets, if they are first class
they are imposed as operatorial identities on the
wavefunctions. Finally, one needs to implement the canonical
transformation as a unitary evolution operator.

\subsection{A cosmological example}
\label{cosmosection}
We will consider a Friedmann cosmological model, written in terms of
Ashtekar's variables \cite{kodama}. The fundamental canonical pair is
$(E,A)$ where $E$ is the only remnant of the triad after the
minisuperspace reduction and $A$ is its canonically conjugate
variable. We will consider the presence of a cosmological constant and
of a scalar field. We will assume the scalar field has a very large
mass so we can neglect its kinetic term in the Hamiltonian constraint,
for the sake of computational simplicity. The Lagrangian for the model
is,

\begin{equation}
L= E\dot{A} + \pi \dot{\phi}- N E^2 (-A^2+(\Lambda +m^2 \phi^2)|E|)\label{lag}
\end{equation}
where $\Lambda$ is the cosmological constant, $m$ is the mass of the
scalar field $\phi$, $\pi$ is its canonically conjugate momentum and
$N$ is the lapse with density weight minus one. The appearance of
$|E|$ in the Lagrangian is due to the fact that the term cubic in $E$
is supposed to represent the spatial volume and therefore should be
positive definite. 

We consider the evolution parameter to be a discrete variable. Then the
Lagrangian becomes
\begin{equation}
L(n,n+1)=E_n (A_{n+1}-A_n)+ \pi_n (\phi_{n+1}-\phi_n) -N_n E_n^2 (-A^2_n+
(\Lambda+m^2 \phi^2_n) |E_n|)
\end{equation}

If one carries out the construction of the previous subsection, one
ends up with the following evolution equations, (see \cite{cosmo} for
details),
\begin{eqnarray}
P^A_{n+1}&=&A_n^2 \Theta^{-1}\label{ecupa}\\
A_{n+1}&=&{3A_n^2-P^A_n\Theta\over 2A_n}\label{ecua}\\
\phi_{n+1}&=&\phi_n\\
P^\phi_{n+1}&=&P^\phi_n -\left(A_n^3-P^A_n\Theta A_n\right)m^2\phi_n\Theta^{-2}
\end{eqnarray}
$\Theta=\Lambda+m^2\phi_n^2$. The variable $E$ and its canonical
momentum have been eliminated, and we have solved the 
constraints of the model  for the Lagrange multiplier $N$. This
evolution preserves the Poisson brackets among the canonical 
pairs $(A,P^A)$ and $(\phi,P^\phi)$. 

The classical solution of the finite difference equations can be seen
to approximate well the continuum solution (see \cite{cosmo} for
details). The approximation gets better for later times, and one 
enters a ``continuous regime'' in which one can approximate the
dynamics by a differential equation in terms of a continuous variable.
Such equation can be integrated exactly and from there we can read
the behavior of the variables asymptotically into the future.
The result is,
\begin{eqnarray}
A&=&\ell_{\rm Planck} \sqrt{\Lambda}a (n+k)^{2/3},\\
E&=&\ell^2_{\rm Planck} a^2 (n+k)^{4/3},
\end{eqnarray}
where $a$ and $k$ are integration constants, which we chose to be
dimensionless and therefore we made explicit the appropriate
dimensions through the introduction of the Planck length $\ell_{\rm
Planck}$.
The connection $A$ is therefore dimensionless since the cosmological 
constant $\Lambda$ has  dimensions of $(\rm length)^{-2}$.

The model can be quantized by implementing the canonical evolution
equations as operator equations via a unitary evolution operator,
\begin{equation}
<A',\phi'|U|A,\phi> = {\rm sg}(A) \sqrt{2 |A| \over \Theta} e^{i{\rm
sg}(A) A^2 {A'-A \over \Theta}} \delta(\phi'-\phi)
\end{equation}

Dirac \cite{Dirac} had already noted in 1933 that the unitary operator
that implements a canonical transformation is given by $\exp(-i G)$
where $G$ is the generating function of the canonical
transformation. In our case the generating function (after eliminating
the Lagrange multipliers is indeed given by $G(A_{n},A_{n+1})=
A^2_{n}(A_{n+1}-A_{n})/\Theta$.  There is an overall difference with
Dirac's result since he chooses a specific factor ordering that does
not coincide with the one we chose. It is interesting that this 
construction is what led Dirac to the notion of path integral.

To define a time we introduce the conditional probabilities,
``probability that a given variable have a certain value when
the variable chosen as time takes a given value''. 
For instance, taking $A$ as our time variable, let us work out first
the probability
that the scalar field conjugate momentum be in the range 
$\Delta P^\phi= [P^\phi_{(1)},P^\phi_{(2)}]$
and ``time'' is in the range $\Delta A=[A_{(1)}, A_{(2)}]$
(the need to work with ranges is because we are dealing with continuous
variables). Since we have no constraints, the
wavefunctions $\Psi[A,\phi,n]$ in the Schr\"odinger representation 
admit a probabilistic interpretation. Therefore the probability 
of simultaneous measurement
is, 
\begin{equation}
P_{\rm sim}(\Delta P^\phi,\Delta A) = \lim_{N\to \infty} {1 \over N}
\sum_{n=0}^N \int_{P^\phi_{(1)},A_{(1)}}^{P^\phi_{(2)},A_{(2)}}
\Psi^2[A,P^\phi,n] dP^\phi dA. \label{psim}
\end{equation}
We have summed over $n$ since there is no information about the ``level''
of the discrete theory at which the measurement is performed, since $n$
is just a parameter with no physical meaning. With the normalizations
chosen if the integral in $P^\phi$ and $A$ 
were in the range $(-\infty,\infty)$, $P_{\rm sim}$ would be equal to one.

To get the conditional probability $P_{\rm cond}(\Delta P^\phi|\Delta A)$,
that is, the probability that having observed $A$ in $\Delta A$ we also
observe $P^\phi$ in $\Delta P^\phi$, we use the standard probabilistic
identity 
\begin{equation}
P_{\rm sim}(\Delta P^\phi,\Delta A) =
P(\Delta A) P_{\rm cond}(\Delta P^\phi|\Delta A)
\end{equation}
where $P(\Delta A)$ is obtained from expression (\ref{psim})
taking the integral on $P^\phi$ from $(-\infty,\infty)$. We therefore
get 
\begin{equation}
P_{\rm cond}(\Delta P^\phi|\Delta A)=
{\lim_{N\to \infty} {1 \over N}
\sum_{n=0}^N \int_{P^\phi_{(1)},A_{(1)}}^{P^\phi_{(2)},A_{(2)}}
\Psi^2[A,P^\phi,n] dP^\phi dA \over
\lim_{N\to \infty} {1 \over N}
\sum_{n=0}^N \int_{-\infty,A_{(1)}}^{\infty,A_{(2)}}
\Psi^2[A,P^\phi,n] dP^\phi dA}.  
\end{equation}
Notice that all the integrals are well defined and the resulting 
quantity behaves as a probability in the sense that integrating 
from $(-\infty,\infty)$ in $P^\phi$ one gets unity.

We will return to this example when we attempt to estimate the effects
of decoherence in our current universe.

\section{Relational time in generally covariant systems}

Let us return now to the general discussion of the use of a relational
time in generally covariant systems. We assume we have discretized the
system of interest using the techniques describe in the previous
section.  Therefore the resulting discrete theory is either free of
constraints, or if, there are constraints left, we need to ensure we
work with physical quantitites that have vanishing Poisson brackets
with the remaining constraints. An example of this would be general
relativity written in terms of Asthekar variables in a generic
situation. There our technique eliminates the diffeomorphism and
Hamiltonian constraint but leaves the Gauss law as a constraint. One
therefore would have to build the arguments that follow involving
quantities that are invariant under gauge transformations, for
instance using Wilson loops.

As we discussed in the previous section, we have a unitary operator to
be related to the canonical transformation on the phase space $q_n,p_n
\rightarrow q_{n+1},p_{n+1}$. We shall denote this operator as
$U(n,n')$ for a general displacement $n \rightarrow n'$. At this point
it is worthwhile re-examining the Pauli argument against promoting
time to a quantum operator. Since in the discrete approach the
evolution of the system from $n$ to $n+1$ is done through a unitary
transformation, corresponding to a discrete canonical transformation
in the classical theory, the evolution is not associated to a
Hamiltonian. One can define (locally in time) objects that behave
close to a Hamiltonian at least when one considers configurations that
approximate the continuuum well. But such objects do not exist
globally \cite{smolin}. Therefore one generically does not have a
Hamiltonian with which to build the argument laid out by Pauli, which
is of global nature. Notice that in terms of the parameter $n$ the
evolution is globally defined, but $n$ is not what is promoted to an
operator representing time.

Let us introduce now the set of self-adjoint operators ${\hat
O}_n({\hat q}_n,{\hat p}_n)$ with eigenvalues $o$ on the isomorphic
set of Hilbert spaces ${\cal H}_n=L^2(q_n)$. We can construct now the
set of projectors ${P_o}(n)=\int_{\Delta o}\sum_j|o,j,n><o,j,n|$,
where we have assume a continuum spectrum, and denote the eigenvalues
of all other operators that form a complete set with ${\hat O}$, as
$j$.

We will construct the conditional probability to measure the value $o$
(since we consider a continuous spectrum, one should strictly speak of
measuring a value within an interval $\Delta o$ surrounding the value
$o$) for the partial observable $O(q_n,p_n)$.  We also consider a
``clock variable'' $T(q,p)$ and the associated self-adjoint operator
${\hat T}_n({\hat q}_n,{\hat p}_n)$ with eigenvalues $t$. We define
the projector $P_t(n)=\int_{\Delta t}\sum_k|t,k,n><t,k,n|$ as we did
for ${\hat O}_n$. In this case $k$ denotes the eigenvalues of all the
operators that form a complete set with ${\hat T}$. These are well
defined projectors for each $n$ value, that is
${P_a(n)}^2=P_a(n),P_a(n)P_{a'}(n)=\delta(a-a')P_a(n),\sum_a
P_a(n)=1\; \forall\; n$.

We are now ready to introduce the relational time. For an application
of these ideas in the simple example of the parameterized non-relativistic particle see
\cite{greece}.  The relational interpretation is defined as
follows. Let us assume that the system is initially in the state
described by the density matrix $\rho$.  The conditional probability
to obtain the value $o \in \Delta o$ for the quantity $O$ given the
value $t \in \Delta t$ for $T$ is,
\begin{equation}
{\mathcal{P}(o\in \Delta o|t \in \Delta t)}_{\rho}=\frac{\sum_n
{\rm Tr}(P_o(n)P_t(n)\rho P_t(n))} {\sum_n {\rm Tr}(
P_t(n)\rho)}.\label{prob1}
\end{equation}

It should be emphasized the important role that the presence of the
parameter $n$ plays in these formulas. Without such an ordering
parameter, one could not define the conditional probabilities.  The
parameter $n$ introduces a notion of simultaneity in the construction
(at a given spatial point).  In fact, several previous attempts to
introduce relational times were problematic due to the lack of such a
parameter. For instance, this led Unruh \cite{unruhmoscow} to attempt
to introduce in an ad-hoc manner a  ``mysterious time'' in the
continuuum theory to play such a role.

If  we further assume that $\hat{O}$ and $\hat{T}$ commute, we
can construct the projector
\begin{equation}
P_{o,t}(n)=\sum_l\int_{\Delta o,\Delta t} |o,t,n,l><o,t,l,n|
\end{equation}
and
rewrite equation (\ref{prob1}) as,
\begin{equation}
{{\cal P}(o \in \Delta o|t \in \Delta t)}_{\rho}=\frac{\sum_n
{\rm Tr}(P_{o,t}(n)\rho)} {\sum_n {\rm Tr}(\int do P_{o,t}(n)\rho)} \label{eq20}
\end{equation}
and from now on for simplicity we will assume that $\hat{O}$ and
$\hat{T}$ commute.

Due to the unitary evolution and the cyclic property of the trace, if
we define the operators $\Pi_o=\sum_n P_o(n)$,$\Pi_t=\sum_n P_t(n)$
and $\Pi_{o,t}=\sum_n P_{o,t}(n)$, we can rewrite the conditional
probability as,
\begin{equation}
{\mathcal{P}(o \in \Delta o|t \in
\Delta t)}_{\rho}=\frac{{\rm Tr}(\Pi_{o,t}\rho)} {{\rm Tr}(\int do
\Pi_{o,t}\rho)} \equiv \frac{{\rm Tr}(\Pi_{o,t}\rho)} {{\rm Tr}(\Pi_{t}\rho)}.
\label{prob2}
\end{equation}

The reduction postulate is given by,
\begin{equation}
\rho \rightarrow \sum_n P_{o,t}(n)\rho P_{o,t}(n).\label{colap}
\end{equation}

This reduction process allows us to calculate the conditional
probability that defines the correlation functions (propagators),
\begin{equation}
{\cal P}(o'|t',o,t,\rho)=\frac{\sum_{n,n'}
{\rm Tr}(P_{o',t'}(n')P_{o,t}(n)\rho P_{o,t}(n))}
{\sum_{n,n'}{\rm Tr}(P_{t'}(n')P_{o,t}(T)\rho P_{o,t}(n))}.
\label{probrel2}
\end{equation}

The difference with usual propagators is that here the times $t,t'$
are the outcome of a quantum measurement. Therefore this will lead to
a meaningful definition of probability if we interpret the above
expression as having measured $o$,$t$ and $t'$ and asking what is the
probability that one will measure $o'$ given those measurements. 
Notice that at the moment there is no well defined concept of
``time ordering'', since the latter is only expected to arise in
semiclassical regimes. There is however a well defined notion of
simultaneity (at a given spatial point). In this sense the
measurements of $o$ and $t$ should be simultaneous and so should
be the measurement of $t'$ and $o'$.

In non-relativistic quantum mechanics one can compute the
probability that measurements of position will find the particle
at a position x at a series of times $t_1, t_2, .... t_n$. Here one
can compute such probabilities through a natural extension of
(\ref{probrel2}). This introduces a reduction postulate in the
framework and leads to the history approach for computing a probability
for a succession of events. This works if one chooses as clock a
robust variable, i.e., such that the information about the
variable  is not destroyed in the  measurements.

From the previous expression, we have that if one prepares
a quantum state with eigenvalue $o$ at time $t$, $\rho_{o,t}$
the probability of it to evolve into a state with eigenvalue $o'$
at time $t'$ can be written as,
\begin{equation}
{{\cal P}(o \in \Delta o'| t \in \Delta t')}_{\rho_{o,t}}=
\frac{Tr(\Pi_{o',t'}\rho_{o,t})}
{Tr(\Pi_{t'}\rho_{o,t})}.\label{prob3}
\end{equation}

Since up to now we have made almost no assumptions about the nature of
the ``time'' $t$ chosen, it could happen that the same value of $t$
occurs many times upon ``evolution'' in the parameter $n$. This
eliminates the predictive power of the theory, at least locally in
time in the following sense: one could make a definite prediction only
upon completing the entire evolution of the system and determining if
the variable $t$ takes a given value more than once. Only then one
could make sense of the probabilities and predict the probability of a
given observable taking a given value at ``time $t$''.  We will
discuss this in detail in a forthcoming paper.

As we shall see in what follows, there is a particular regime where we
can concentrate on a particular range of steps. This regime will
correspond to considering as a clock a variable that operates 
semiclassically. When this happens, then the quantum mechanics
we constructed will reproduce the results of ordinary quantum
mechanics. The agreement will be better the more classically
the clock variable behaves.

Another element to be emphasized is that we have used the parameter
$n$ to define simultaneity. One has to be careful in the case of
general relativity that one has discretized the theory in such a way
that the parameter $n$ is associated with Cauchy surfaces of the
continuum theory. Otherwise it would not be correct to use the
parameter to define a notion of simultaneity. Even this requirement is
not enough. In the consistent discretization scheme the lapse is
determined dynamically. One cannot rule out situations in which the
lapse becomes negative. In such situations ``time runs backwards'' and
one covers the same region of the manifold more than once. These 
situations in the classical theory should also be avoided in order
to obtain a sensible quantum mechanics. It should also be emphasized
that the whole issue of the covariance of the formalism is still to be
worked out in detail. In particular, since we are here largely
concentrating in models without spatial degrees of freedom, it is
difficult to even pose the question of how they transform under
coordinate transformations. We are taking into account possible time
reparameterizations, but not coordinate changes that mix space and
time. We are currently studying model systems in which some limited
set of symmetries can be implemented and study how the relational
approach will mix with covariance, but we will not discuss this in
this paper.

\section{Semiclassical time}

We now wish to show that the new quantum mechanics we have created
reduces to ordinary quantum mechanics if the time variable chosen
behaves classically. If one takes into account semiclassical
corrections for the time variable, then one will end up with
corrections to ordinary quantum mechanics. At this point it is
worthwhile mentioning the work of Egusquiza, Garay and Raya
\cite{EgGaRa}. They have studied modifications to quantum mechanics
through the use of imperfect clocks. They treat the clocks classically
but admit they may have fluctuations in their behavior, perhaps of
quantum mechanical origin, and they model them through a Markovian
process with a given probability distribution. There are many points
of contact between their calculations and the ones we present
here. The main difference relies on the fact that we are treating the
clock as quantum mechanical and in particular we allow its
probability to evolve as a function of time. We are not including any
thermal or other effect, but these could definitely be incorporated if
needed through the formalism of Egusquiza {\rm et al.}

Let us assume now the existence of a semiclassical regime for the 
variable chosen as time for a given initial state of the complete
system $\rho$. That means that there exists a region $R$ 
in the spectrum of the operators ${\hat
T}_n$ such that for a value $t\in R$ there is an interval
of values $\Delta_tn$ of the step parameter $n$
such that 
${\cal P}_n(t) \approx 0 \forall n\notin {\Delta_tn}$. Also
given a value of the step parameter $n_0$, there exists an
interval around it $\Delta_{n_0}t$ such that 
${\cal P}_{n_0}({\tilde t})\approx 0 \;
\forall \; {\tilde t} \notin \Delta_{n_0}t=0$. In these expressions ${\cal
P}_n(t)$ is the probability density of having the value $t$ at a
given $n$. We speak of probability density since we are assuming that the
operator $T$ has a continuous spectrum. If the spectrum were discrete then
${\cal P}_n(t)$ would be the probability of having the value $t$ at the
level $n$. This semiclassical limit therefore implies a 
strong correlation between the $n$ parameter and the eigenvalues of the
observable $T$ in region $R$. Notice that
in this case each reduction process is almost a projection given
by,
\begin{equation}
\rho \rightarrow \sum_{n\in\Delta_tn} P_{o,t}(n)\rho P_{o,t}(n).
\end{equation}

We need to assume also that the clock and the system in study
are weakly interacting (otherwise one would not have a reasonable
clock). Concretely, we assume that the state of the system is of the
form of a product
$\rho \approx
\rho_1\bigotimes \rho_2$ throughout the evolution determined by a
unitary operator $U\approx U_1\bigotimes U_2$ also of product form. 
We take ``1'' to describe the clock and ``2'' the physical system
under study and the weak interaction also implies that $U_1$ and
$U_2$ commute.

Up to now we have considered the quantum states as described by a
density matrix at a given level of the discretization parameter $n$.
Since the latter is not an observable, we would like to transition to
a description where we have density matrices function of the
observable time instead of $n$. In order to do this, let us recall the
expression for the usual probability in the Schr\"odinger
representation of measuring the value $o$ at time $t$ in ordinary
quantum mechanics,
\begin{equation}
{\cal P}(o|t)_\rho \equiv 
\frac{{\rm
Tr}(P_{o}(0)\tilde{\rho}(t))}{{\rm Tr}(\tilde{\rho}(t))},\label{usual}
\end{equation}
where the projector is evaluated at $t=0$ since in the Schr\"odinger
representation the operators do not evolve. We would like to obtain
an expression similar to this one in the relational time picture.
We are denoting $\tilde{\rho}(t)$ as the density matrix in the 
picture in which we have a time variable, we will drop the tilde to denote
the density matrix that arises under the evolution of the 
step parameter $n$ and we will drop the $n$ dependence to 
refer to the initial density matrix  $\rho(n=0)$.

We start by considering the conditional probability defined in (\ref{eq20}),
\begin{equation}
{\cal P}(o \in \Delta o | t \in \Delta t)_\rho = 
{ \sum_{n} {\rm Tr}\left( P_o(n) P_t(n) \rho P_t(n)\right) \over
\sum_n {\rm Tr}\left(P_t(n) \rho\right)},
\end{equation}
and one could have omitted the last $P_t(n)$ using the cyclicity of 
the trace since we are assuming that $P_t$ and $P_o$ commute. Given
that we are assuming the presence of continuous spectra, the
projectors in the above expression should be understood as integrated
over the interval, i.e. $P_o(n) = \int_{\Delta o} P_{o'}(n) do'$
and similar for $P_t$. We now introduce the hypothesis that the
clock and the rest of the system interact weakly and write explicitly
the evolution of the projectors in the step parameter $n$ to get,
\begin{eqnarray}
{\cal P}(o \in \Delta o | t \in \Delta t)_\rho &=& {
\sum_n {\rm Tr}\left(U^\dagger_2(n) P_o(0) U_2(n) U^\dagger_1(n) P_t(0) U_1(n) \rho_1 \otimes \rho_2\right)
\over \sum_n {\rm Tr}\left(P_t(n) \rho_1\right) {\rm Tr}\left(\rho_2\right)}\nonumber\\
&=& {
\sum_n {\rm Tr}\left(U^\dagger_2(n) P_o(0) U_2(n)\rho_2\right){\rm Tr}\left( 
U^\dagger_1(n) P_t(0) U_1(n) \rho_1 \right)
\over \sum_n {\rm Tr}\left(P_t(n) \rho_1\right) {\rm Tr}\left(\rho_2\right)}.\label{condprob}
\end{eqnarray}

From this expression, using the cyclic property of the trace, we can identify
the expressions of the density matrix evolved in relational time. We start by
defining the probability that the measurement $t$ corresponds to the value $n$,
\begin{equation}
{\cal P}_n(t) \equiv {{\rm Tr}\left( 
P_t(0) U_1(n) \rho_1 U^\dagger_1(n) \right)
\over \sum_n
{\rm Tr}\left( 
P_t(n)\rho_1  \right)},
\end{equation}
and notice that $\sum_n {\cal P}_n(t)=1$.

We now define the evolution of the density matrix,
\begin{eqnarray}
\tilde{\rho}_2(t)\equiv \sum_n U_2(n) \rho_2 U^\dagger_2(n) {\cal
P}_n(t), \label{evo1}
\end{eqnarray}
and noting that 
\begin{equation}
{\rm Tr}\left(\tilde{\rho}_2(t)\right) 
= \sum_n {\cal P}_n(t) {\rm Tr}\left(\rho_2\right)={\rm Tr}(\rho_2)
\end{equation}
one can equate the conditional probability (\ref{condprob}) with the usual
expression for a probability in quantum mechanics (\ref{usual}).
It should be noted that all the sums in $n$, due to the assumption that the
time variable is semiclassical are only nontrivial in the interval $\Delta_t n$ 
since outside of it, probabilities vanish. Something else to notice is that 
when we introduced the projectors, there was an integral over an interval.
Therefore in the above expression for the evolution of the density 
matrix, this has to be taken into account. Since the interval $\Delta t$ is
arbitrary, one can consider the limit in which its width tends to zero,
apply the mean value theorem in the integrals, and the interval in the
numerator and denominator cancel out, yielding an expression for 
$\tilde{\rho}_2(t)$ that is independent of the interval, and involves 
the non-integrated projector  $P_t(0)$.

We have therefore ended with the standard probability expression with
an ``effective'' density matrix in the Schr\"odinger picture given by
$\tilde{\rho}_2(t)$.  In its definition, it is evident that unitarity
is lost, since one ends up with a statistical mixture of states
associated with different $n$'s.  We also notice that probabilities
are conserved, as can be seen by taking (\ref{usual}) and integrating
over $x$. We recall that $\tilde{\rho}_2$ is not the normalized
density matrix; the latter can be easily recovered dividing by the
trace.

We will assume that ${\cal P}_n(t) \equiv f(t-t_{max}(n))$ with
$f$ a function that decays quite rapidly for values of $t$ distant of
the maximum $t_{\rm max}$ which depends on $n$. 

To manipulate expression (\ref{evo1}) more clearly, we will assume we
are considering a finite region of evolution and we are in the limit
in which the number of steps in that region is very large. We denote
the interval in the step variable $n$ as going from zero to $N$ with
$N$ a very large number. We define a new variable $v=\epsilon n$ with
dimensions of time such that $N \epsilon=V$ with $V$ a chosen finite
value. We can then approximate expression (\ref{evo1}) by a continuous
expression,
\begin{equation}
\tilde{\rho}_2(t)= \int_0^V dv f(t-t_{max}(v)) \rho_2(v).
\label{evocont}
\end{equation}

In this expression $t_{\rm max}(v)\equiv t_{\rm max}(n=v/\epsilon)$ and 
\begin{equation}
\rho_2(v) = U_2(n=v/\epsilon) \rho_2 U^\dagger_2(n=v/\epsilon).
\end{equation}
In all the above expressions, when we equate $n=v/\epsilon$ it should
be understood as $n={\rm Int}(v/\epsilon)$, which coincide in the
continuum limit. (Notice that strictly speaking we should write $\rho_2(v/\epsilon)$
to keep the same functional form as for $\rho_2(n)$, but we will drop the 
$\epsilon$ to simplify the notation.)

Let us assume now that $t_{max}(v)=v + \epsilon \Gamma_1(v)+
\epsilon^2 \Gamma_2(v)$, with $\Gamma_1(v)\sim 1$ and $\Gamma_2(v)\sim
1$, that is, the value at which ${\cal P}_n(t)$ has its maximum
depends linearly on $n$ plus a small correction. One can always obtain
the classical dependence of $t$ on $n$ solving the equations of the
theory.  The linear approximation can be justified locally as a Taylor
expansion, or by redefining the variable chosen as time such that the
relation with $n$ is linear.

To simplify calculations we replace the function $f$ with a function $\tilde{f}$
which agrees with it up to terms of order $\epsilon^2$,
\begin{equation}
f(t-t_{\rm max}(v))=f(t-v-\epsilon \Gamma_1(v)-\epsilon^2\Gamma_2(v))\equiv
\tilde{f}(v-t+\epsilon \Gamma_1(t)+\epsilon^2 \Gamma_2(t)
-\epsilon^2 \alpha \Gamma_1(t)),
\end{equation}
where
$\alpha(v)\equiv \partial \Gamma_1(v)/\partial v$ and we have neglected terms
of order $\epsilon^3$.

We now assume that $f$ is well approximated by a
Dirac delta,
\begin{eqnarray}
\tilde{f}(v-t+\epsilon \Gamma_1(t)+\epsilon^2 \Gamma_2(t)
-\epsilon^2 \alpha \Gamma_1(t))&=&c_0(t)\delta(v-t+\epsilon \Gamma_1(t)+\epsilon^2 \Gamma_2(t)
-\epsilon^2 \alpha \Gamma_1(t))\nonumber\\ &&+ a(t) \epsilon \delta'
(v-t+\epsilon \Gamma_1(t)) + b(t) 
\epsilon^2 \delta''(v-t)+O(\epsilon^3) 
\label{deltadelta}
\end{eqnarray}
with $b>0$. Since ${\cal P}_v(t)=\tilde{f}(v-t+\epsilon
\Gamma_1(t)+\epsilon^2 \Gamma_2(t) -\epsilon^2 \alpha \Gamma_1(t))$ and 
since we want $\int_0^V {\cal P}_v(t)=1$ then $c_0(t)=1$.

We now substitute the explicit form of $\tilde{f}$ (\ref{deltadelta})
in (\ref{evocont}), which allows to compute the integral in $v$
explicitly,
\begin{equation}
\tilde{\rho}_2(t)=\rho_2(t-\epsilon\Gamma_1(t)-\epsilon^2\Gamma_2(t)
+\epsilon^2 \alpha(t) \Gamma_1(t))-a(t) 
\epsilon {\partial \over \partial v} \rho_2(t-\epsilon\Gamma_1(t))+
b(t) \epsilon^2 {\partial^2 \over \partial v^2} \rho_2(t)+O(\epsilon^3).
\end{equation}
 
We proceed to expand the arguments of $\rho_2$ in this expression in
powers of $\epsilon$. We also use the fact that in the continuum limit
we can now associate a Hamiltonian with the evolution operator
$U_2(v)=\exp(i H_2 v)$, and  recalling that the evolution of the density
matrix is $\rho_2(v)=U_2(v)\rho_2 U^\dagger_2(v)$, we get,
\begin{eqnarray}
\tilde{\rho}_2(t)&=&\rho_2(t)+ (i\epsilon(\Gamma_1(t)+a(t)))[H_2,\rho_2(t)]
-i\epsilon^2\left(\alpha(t) \Gamma_1(t)-\Gamma_2(t)\right) 
[H_2,\rho_2(t)]\nonumber\\&&-\epsilon^2 
({1 \over 2} \Gamma_1(t)^2+a(t)\Gamma_1(t)+b(t)) 
[H_2,[H_2,\rho_2(t)]].\label{39}
\end{eqnarray}
Notice that generically, one can write the discrete generator as
$U_2(n)=\exp(i H^{\rm dis}_2 n)$. However, there may be points where the
canonical transformation is singular, and example is the Big Bang in
the cosmological model \cite{smolin}, the series defining the
logarithnm of $U_2$ fails to converge and therefore $H_2$ does not
exist. Notice also that in taking the continuum limit $H^{\rm
dis}_2/\epsilon=H_2$.

We will find useful to have the inverse relation between the density
matrices,
\begin{eqnarray}
\rho_2(t)&=&\tilde{\rho}_2(t)- (i\epsilon(\Gamma_1(t)+a(t)))[H_2,\tilde{\rho}_2(t)]
- \epsilon^2 (\Gamma_1(t)+a(t))^2 [H_2,[H_2,\tilde{\rho}_2(t)]]\label{38}\\
&&+i\epsilon^2\left(\alpha(t) \Gamma_1(t)-\Gamma_2(t)\right)  [H_2,\tilde{\rho}_2(t)]
+\epsilon^2  \left({1 \over 2} \Gamma_1(t)^2+a(t)\Gamma_1(t)+b(t)\right) 
[H_2,[H_2,\tilde{\rho}_2(t)]].\nonumber
\end{eqnarray}

We now take the time derivative of equation (\ref{39}) and we note 
that the derivatives will act on the coefficients
$a,b,\alpha,\Gamma(t)$ or on $\rho_2(t)$.  We replace
$\dot{\rho}_2(t)$ with
the commutator with the Hamiltonian. The final result, using
(\ref{38}) to rewrite everything in terms of 
$\tilde{\rho}_2$ is
\begin{equation}
{\partial \tilde{\rho}_2(t) \over \partial t} = 
i \left(-1+\epsilon{\partial \over \partial t} \left({\Gamma}_1(t)+{a}(t)\right)
-\epsilon^2 {\partial (\alpha(t)\Gamma_1(t)-\Gamma_2(t)) \over \partial t} 
\right)[H_2,\tilde{\rho}_2(t)]+\epsilon^2 \left[{\partial \over
\partial t} 
\left({a(t)^2\over 2} -b(t)\right)\right] [H_2,[H_2,\tilde{\rho}_2(t)]].
\end{equation}

The general form of the resulting evolution equation for the system under
study is therefore,
\begin{equation}
\frac {\partial}{\partial t}{\tilde \rho_2}=-i[(1+\beta(t))
{H}_2,{\tilde \rho_2}]-\sigma(t)[{H}_2,[{H}_2,{\tilde
\rho_2}]],\label{ec2}
\end{equation}
with the coefficients $\beta(t)$ and $\sigma(t)$ that are functions
that are small corrections (in terms of the expansion in $\epsilon$),
and we have neglected terms involving triple and higher commutators
with the Hamiltonian since they correspond to higher powers of $\epsilon$

The presence of $\beta(t)$ just redefines the Hamiltonian of the theory.
The coefficient $\sigma(t)$ is the one that gives rise to new effects. In
particular it will imply that pure states
evolve into mixed states. In fact, the equation we have obtained is a particular
case of equations that are considered in the context of decoherence in 
quantum mechanics, called Lindblad \cite{lind} type equations
\begin{equation}
\frac{\partial\rho}{\partial t}=-i[H,\rho]-{\cal
D}(\rho)\label{rho},
\end{equation}
with ${\cal D}(\rho)$ satisfying the properties,
\begin{equation}
{\cal D}(\rho)=\sum_n[D_n,[D_n,\rho]], \;\;\;
D_n=D_n^{\dagger},\;\;\; [D_n,H]=0,
\end{equation}
so it defines a completely positive map on $\rho$ that  is
consistent with the monotonous increase of Von Neumann entropy $S={\rm
Tr}(\rho \log \rho)$ and conservation of energy. This type of equation
was introduced by Ghirardi, Rimini and Weber (GRW) \cite{ghir} with
the aim of providing an objective solution to the measurement problem
in standard quantum mechanics. (Similar equations can be used to
describe the decoherence due to interaction with an environment, see
\cite{joos}.) GRW considered a single $D$ as a localizing operator in
coordinate space. As discussed by Adler and Horwitz \cite{adler}, and
also Milburn, Percival and Hughston
\cite{mil,perc}, setting $D$ to be proportional to 
$H$ is most natural since it leads to an objective state vector
reduction in the energy pointer basis. This loss of coherence may be a
way to avoid macroscopic superpositions, like the ``Schr\"odinger
cat'' \cite{mil,adler}.  

We can recognize that in our case there is only one $D_n$ that is
given by the Hamiltonian. Having a Lindblad form is desirable since it
implies that the Hamiltonian will be conserved automatically by the
evolution considered.  Other proposals for decoherence from quantum
gravity, like Hawking's \$-matrix may have problems with the
conservation of energy \cite{hag}.

In order to study the influence of $\beta(t)$ and $\sigma(t)$, we
start by assuming that they are constant as functions of time
$\beta(t)=\bar{\beta}$ $\sigma(t)=\bar{\sigma}$. In general these
functions will be constant will some small fluctuations, in which case
we can interpret the bars as average values.  In such case the
equation can be solved exactly,
\begin{equation}
{\rho_2}_{nm}(t)= {\rho_2}_{nm}(0)e^{-i(1+{\bar
\beta})\omega_{nm}t} e^{(-{\bar
\sigma}(\omega_{nm})^2)t},
\end{equation}
where we have written the density matrix element in the basis of
energy eigenstates, and $\omega_{nm}$ is the Bohr transition frequency
between the state $n$ and $m$ in the energy basis. It is important 
to notice that $\bar{\sigma}$ has to be positive for the evolution to
be physically acceptable. Otherwise the trace of the square of the
density matrix will be larger than one.

Let us estimate the sign of 
\begin{equation}
\sigma(t)=-\epsilon^2 \left[{\partial \over
\partial t}  \left({a(t)^2\over 2} -b(t)\right)\right]. 
\end{equation}
The coefficient $a(t)$ represents the asymmetry of the probability of
the time $t$ as a function of the evolution parameter $n$. There is no
reason for this probability distribution to have a definitive
asymmetry, so its mean value through the evolution will vanish.
The coefficient $b(t)$ represents the spread of the probability
distribution. Suppose one starts the clock in a quantum state
that is peaked around a certain value $t_0$. As the parameter $n$
evolves, the quantum state will disperse and the spread of the
probability of $t$ as a function of $n$ will disperse too. Therefore
$b(t)$ will increase and its derivative will be positive. The 
magnitude of the effect will depend on the details of the clock.
For instance, if one used a free particle as clock the spread of
the wavepacket will be linear and therefore $\sigma(t)$ will be a
constant. The important point is that it is positive, and therefore
signals an arrow of time.

In the real world, the subsystem chosen as a clock will be subject to
several other kinds of fluctuations, for instance thermal
fluctuations. The influence of these kinds of errors 
leads \cite{EgGaRa}  to an equation similar to (\ref{ec2}), but
where the $\sigma(t)$ is proportional to $\alpha^2$. We therefore see
that a quantum evolution will exhibit decoherence from fundamental
effects, like the ones considered in this paper, and also environmental
effects. The latter could in some cases be minimized by choosing the 
experimental setup in appropriate ways, the fundamental effects will,
however, be an ultimate limit to the achievable coherence in a quantum
evolution.

It should be noted that the value of $\sigma$ depends the choice of 
clock, and on the quantum state chosen for the clock initially, and
on the interaction of the clock with the environment.

\section{Estimating the magnitude of the decoherence}

We will now proceed to use the cosmology we discussed earlier to give
an estimate for the level of decoherence. This can only be viewed as a
first calculation, a more realistic model would be desirable. In
particular, we have chosen a model with a scalar field in the infinite
mass limit, in which the model is a DeSitter universe. This is not a
very good model for approximating the universe since it is homogeneous
in time (and in particular the ``Big Bang'' we refer to is just a
coordinate singularity). One should think of it only as asymptotically
approximating our universe into the future. It should only be taken as
a guide of the way to compute $\sigma$. The calculations could be
repeated in a more realistic model, but would be more involved.

The task at hand is to provide a justification for the lattice spacing
of the discrete model. It is evident that such spacing could not be
determined if we just considered a non-gravitational system as our
only object of study. Considering the influence of gravity will allow
us to introduce a fundamental length scale in the problem, the Planck
scale.

Recalling the definition of $b$ in (\ref{deltadelta}) we can relate it
to the quadratic deviation of the time variable from its linear
behavior $t=\epsilon n$, due to the spread as a function of time of
the wavepacket representing the clock through
\begin{equation}
(\Delta t)^2 \equiv \epsilon^2 b = 
{(\Delta \epsilon)^2 \over \epsilon^2} n^2
\epsilon^2
={(\Delta \epsilon)^2 \over \epsilon^2} t^2
\end{equation}
therefore we can estimate $\sigma$ to be,
\begin{equation}
\sigma=\epsilon^2 \dot{b}\sim{(\Delta \epsilon)^2 \over \epsilon^2} t =
{(\Delta \epsilon)^2 \over \epsilon} n
\end{equation}
(we will neglect factors of order one in working out the estimates).

We now need to choose a time variable in the cosmology. As before, we 
will choose it related to the connection $A$, but we also wish to have
a linear relation between time and $n$, as we assumed in the
calculation of the decoherence. Therefore we rescale 
\begin{equation}
t={1 \over \Lambda^{3/4} \ell_{\rm Planck}^{1/2}} A^{3/2},\label{timen}
\end{equation}
which taking into account the asymptotic behavior of the cosmological 
variables we described in section (\ref{cosmosection}) yields,
\begin{equation}
t=a^{3/2} \ell_{\rm Planck} (n+k)
\end{equation}
from which we can therefore read off $\epsilon=a^{3/2} \ell_{\rm
Planck}$
and therefore $\Delta \epsilon = \frac{3}{2} \sqrt{a} \Delta a
\ell_{\rm Planck}$. We are choosing the integration constant $k$ to be
very 
small compared to
$n$, since we are assuming we are in the asymptotic future region.
It should be noted that the singularity happens for $n+k\sim 0$.

So we can therefore estimate $\sigma$,
\begin{equation}
\sigma ={\ell^2_{\rm Planck}a (\Delta a)^2 \over \epsilon} n.
\end{equation}

We can now proceed to find a bound on $a (\Delta a)^2$ based on the
uncertainty principle, $\Delta E \Delta A >\ell^2_{\rm Planck}$, which
yields,
\begin{equation}
a (\Delta a)^2 > { 1 \over \ell_{\rm Planck} \sqrt{\Lambda} n^2}
\end{equation}
so we finally get for the decoherence parameter,
\begin{equation}
\sigma> {\ell_{\rm Planck} \over \epsilon \sqrt{\Lambda} n}
={\ell_{\rm Planck} \over\sqrt{\Lambda}  t}.
\end{equation}

At this point it is worthwhile pondering what is meant in this
equation by ``$t$''. We have identified this variable as related
linearly to the parameter $n$ via equation (\ref{timen}). This 
choice involves selecting a prefactor with dimensions of time.
This prefactor is arbitrary. Therefore it would appear that the result
for $\sigma$ is dependent on this factor. Indeed it is. However, it
should be noted that a similar factor arises in identifying the
continuum limit Hamiltonian $H_2$ from the discrete one $H_2^{\rm
dis}$ as we discussed before. Therefore if one defines a new time
$t'=\lambda t$, the evolution equation for the density matrix acquires
an overall factor $1/\lambda$. This requires rescaling $\sigma$ in the
term involving two Hamiltonians by a factor of $\lambda$ to compensate
for the fact that the Hamiltonian is rescaled by
$1/\lambda$. Therefore although the value of $\sigma$ has this
arbitrariness, the speed at which states decohere is the same, just
measured in different time units.

We still have to ask ourselves how do we connect the time $t$
appearing in the formula for $\sigma$ and the physical time one would
measure in a laboratory experiment. We notice that there exists a
relation between $t$ and $t_{\rm lab}$ of the form $t=t(A(t_{\rm
lab})$. That is, $t$ is connected with the variable we chose as clock
$A$, and the latter is a physical variable of the cosmology that
should be connected with the clock in the lab (just like one measures
the age of the universe in seconds as defined by a regular clock).
This will rescale the left hand side of the evolution equation of the
density matrix of the form 
\begin{equation}
{\partial t \over \partial t_{\rm lab}}
{\partial \rho \over \partial t}
\end{equation}
The prefactor ${\partial t /
\partial t_{\rm lab}}$ will be slowly varying with respect to the time
scales of a laboratory experiment. Therefore we can take it to be a
constant $\lambda$ of the same nature as the one we discussed in the
previous paragraph. This leads to a definition of the $\sigma$ one
would measure in a lab,
\begin{equation}
\sigma_{\rm lab} ={\partial t_{\rm lab} \over \partial t} \sigma.
\end{equation}

In the model we are discussing, the closest thing to a ``physical''
time could be the time of comoving observers. This time has an
exponential relation to the variable $t$ we have been using.
Unfortunately, due to the simplicity of this model there is
an overall scale ambiguity and the relation between the comoving time
and $t$ is defined up to an overall constant $\kappa$. The relation is
$t=\kappa \exp(3\sqrt{\Lambda}t/2)$. Therefore the result for the physical 
$\sigma$ is,
\begin{equation}
\sigma> 
{\ell_{\rm Planck} \over {\Lambda}  \kappa^2 \exp(3 \sqrt{\Lambda} t_{lab})}.
\end{equation}

If one were dealing with a universe that is not exactly DeSitter
but with a Friedmann model with a power law behavior in comoving time,
the relation between comoving time and $t$ would not be exponential
but be of the form $\sqrt{\Lambda}t=({\sqrt{\Lambda}t_{\rm lab}})^\alpha$ with
$\alpha$ some power, since the Big Bang would have to happen at
$t=t_{\rm lab}=0$. If that is the case and we assume one is making an
experiment today, we have that $\sqrt{\Lambda} t_{\rm lab}\sim 1$ and
therefore (we are setting the speed of light $c=1$) one has that
$\sigma \sim t_{\rm Planck}$. As expected, the decoherence effect due
to the finiteness of the (space)-time lattice is of the order of the
Planck scale.

\section{Conclusions}

We have shown how the use of a relational time in quantum mechanics
leads to a modification of ordinary quantum mechanics in which pure
states evolve to mixed states. The use of a relational time is now
possible in quantum gravity due to the use of the consistent
discretization framework. One can then proceed to work out estimates
of the magnitude of fundamental decoherence due to the fact that
in a cosmology time has necessarily to be relational. We have worked
out such estimate in this paper using a simple cosmological model.
It is clear that more detailed models will be needed to gain 
confidence in the estimate obtained.

What are the possibilities of detecting the decoherence due to 
quantum gravity? We have found that the largest amount of decoherence
one can expect in a system goes as 
\begin{equation}
(\omega_1-\omega_2)^2 t_{\rm Planck} t_{\rm life}
\end{equation}
where the $\omega$'s are the Bohr frequencies corresponding to the two
most separated energy levels of the system and $t_{\rm life}$ is how
long we wait for the decoherence to appear. For ordinary quantum
systems such effect is very small. It is interesting to notice
however, that a small level of fundamental decoherence has always
been desirable in quantum mechanics to avoid ``Schr\"odinger cat''
type situations, and it had been advocated through ad-hoc proposals
like that of Ghirardi, Rimini and Weber \cite{ghir} and others.
The most promising experiments where one could observe this effect are
the ones involving macroscopic quantum systems, like Bose--Einstein
condensates. We have discussed some of these possibilities in 
reference \cite{deco1}, but it appears detectability is not within
reach of current technology, largely because the systems involve
a limited number of atoms. Since the energy spread goes as the
number of atoms squared, it might be possible that in a few years
experiments could reach the desired levels of energy spread in
macroscopic quantum systems. It remains to be seen if the specific
experimental setups will allow the effect to be sufficiently isolated
from other sources of decoherence due to environmental effects.

It should be noted that we are claiming that the consistent
discretizations allow to solve the problem of time in quantum gravity
and totally constrained systems in general through the introduction of
a relational time.  The problem of time has many aspects to it. One of
these aspects is the identification of a variable that behaves as a
``good time''. In particular, a variable that is ``transverse'' to the
dynamical orbits of the system. H\'aj\'{\i}\v{c}ek et al. \cite{Haji},
Hartle \cite{Har} and Kucha\v{r} \cite{Ku} have emphasized that in
many cosmological situations there do not exist time variables
transverse to the dynamical orbits. Our approach actually addresses
this problem. In a nutshell, even in such cases the parameter $n$
in the discretization is transverse to the orbits and this is enough
to define a correct relational time. This in particular implies that
our approach solves the ``time of arrival'' problem in quantum
mechanics. We will expand on these issues in a forthcoming
publication.

Another aspect of the framework that requires further analysis is the
issue of the covariance of the predictions. As worked out in this
paper, the predictions are particular to a given choice of time. This
is unsatisfactory since one should expect the theory at least to
exhibit Lorentz invariance locally since it is derived from general
relativity. Establishing this in general relativity is however,
delicate, since it requires analyzing situations with spatial (field
theoretic) degrees of freedom like Gowdy cosmologies which are
considerably more involved computationally.  We are studying
these but we do not have results to report at present. An alternative
would be to consider model situations, for instance, two parameterized
relativistic particles with a relativistic interaction, but this also
requires further study.

Summarizing, the use of consistent discretizations of general
relativity free the theory from constraints and therefore one can
introduce a relational time as proposed by Page and Wootters avoiding
the main objections to such approach. The resulting modified quantum
mechanics implies that pure states evolve into mixed states. We have
computed explicitly the rate for such decoherence. It may play a role
in foundations of quantum mechanics, in solving the black hole
information problem \cite{info}, and may be observed experimentally.

\section{Acknowledgments}
This work was supported by grant NSF-PHY0090091 and funds from the
Horace Hearne Jr. Laboratory for Theoretical Physics.

\end{document}